\documentclass{elsart}

\usepackage{natbib}

\usepackage{epsfig}

\begin{document}

\def\msun{M_\odot}

\begin{frontmatter}



\title{Precision Cosmology}


\author{Joel R. Primack}

\address{Physics Department, 
University of California, Santa Cruz, CA 95064 USA}

\begin{abstract}
The good agreement between large-scale observations and the
predictions of the now-standard $\Lambda$CDM theory gives us hope that
this will become a lasting foundation for cosmology.  After briefly
reviewing the current status of the key cosmological parameters, I
summarize several of the main areas of possible disagreement between
theory and observation: big bang nucleosynthesis, galaxy centers,
dark matter substructure, and angular momentum, updating my earlier
reviews [1].  The issues in all of these are sufficiently complicated
that it is not yet clear how serious they are, but there is at least
some reason to think that the problems will be resolved through a
deeper understanding of the complicated astrophysics involved in such
processes as gas cooling, star formation, and feedback from supernovae
and AGN.  Meanwhile, searches for dark matter are dramatically
improving in sensitivity, and gamma rays from dark matter annihilation
at the galactic center may have been detected by H.E.S.S.
\end{abstract}


\end{frontmatter}

\section{Introduction}

Modern cosmology -- the study of the universe as a whole -- is
undergoing a scientific revolution. New ground- and space-based
telescopes can now observe every bright galaxy in the universe.  We
can see back in time to the cosmic dark ages before galaxies formed
and read the history of the early universe in the ripples of heat
radiation still arriving from the Big Bang.  We now know that
everything that we can see makes up only about half a percent of the
cosmic density, and that most of the universe is made of invisible
stuff called ``dark matter'' and ``dark energy.''  The cold dark
matter (CDM) theory based on this ($\Lambda$CDM) appears to be able to
account for many features of the observable universe, including the
heat radiation and the large scale distribution of galaxies, although
there are possible problems understanding some details of the
structure of galaxies.  Modern cosmology is developing humanity's
first story of the origin and nature of the universe that might
actually be {\it true} -- in the sense that it will still be true in a
thousand years.  Although this talk is entitled ``Precision
Cosmology,'' I think we should be even more impressed that modern
cosmology is true than that it is precise.

Building on the work of Copernicus, Brahe, Kepler, and Galileo, Newton
established the basis of what we now call classical physics.  Although
there have been many scientific revolutions in physics since the
Newtonian synthesis, none of them have overthrown Newtonian physics
the way that the Copernican-Newtonian scientific revolution overthrew
earlier Aristotelian and Ptolemaic ideas.  Ptolemy was never afterward
taught as science, only as history, but Newtonian physics will always
be taught.  The subsequent revolutions in physics -- wave optics,
field theory, thermodynamics, relativity, and quantum mechanics --
encompassed Newtonian physics rather than overthrowing it [2].

Once a well-confirmed basis for further progress is established, such
as that provided by Newtonian physics, a scientific field can expand
its range of successful applicability without any further overthrowing
revolutions, and in this sense it can be said to be progressive.  I think
it is likely that the modern revolution in cosmology has now
established such a basis for progress.  Even though there is so much
that we still do not know -- in particular, the nature of the dark
matter and dark energy, and the origin of the initial conditions --
what we do know is now so well confirmed by diverse data that it is
likely to be true.

Ever since Einstein's general relativity provided the essential
language for cosmology, the field has progressed in the normal
scientific style, with predictions followed by confirmations.
Friedmann and Lemaitre predicted the expansion of the universe, which
was subsequently confirmed by Hubble in 1929.  Gamow, Alpher, and
Hermann in 1948 predicted the existence of the cosmic background
radiation (CBR) which was found by Penzias and Wilson in 1965, with
its thermal spectrum confirmed by the FIRAS instrument on the COBE
satellite in 1989.  The cold dark matter theory [3] predicted the
amplitude of the CBR fluctuations, which were discovered by the DMR
instrument on COBE in 1992 and found to have the predicted amplitude.
By early 1992, it was clear that the only viable simple versions of
CDM were $\Lambda$CDM with $\Omega_m \approx 0.3$ and $\Omega_\Lambda
\approx 0.7$, and Cold+Hot DM (CHDM) with $\Omega_m = 1$ and
$\Omega_\nu = 0.2-0.3$.  A few years later CHDM and all other
$\Omega_m = 1$ cosmologies were ruled out by the discovery of abundant
high redshift galaxies and by the discovery of strong evidence for
$\Lambda$ using high-redshift supernovae in 1998.  The combination of
CDM and cosmic inflation predicted the acoustic peak in the CBR
angular power spectrum, which was discovered by the BOOMERANG and
MAXIMA balloon experiments and the DASI instrument at the South Pole
in 2000-2002.  Now WMAP has confirmed and extended the CBR
observations of ground- and balloon-based instruments, and both the
CBR angular power spectrum and the galaxy power spectrum look exactly
like the predictions of $\Lambda$CDM [4].

Our modern cosmological synthesis is based on several assumptions of
simplicity, in particular the ``cosmological principle'' (we don't
live at a special place in the universe) and the assumption that the
same laws of physics that describe phenomena in our laboratories on
earth and nearby are valid at all times and places throughout the
universe.  These assumptions are being checked against observations.
For example, comparison of the details of atomic spectra in the
laboratory and from galaxies at various redshifts suggested that there
might be variations in the fine structure constant $\alpha =
e^2/(\hbar c)$ of $\Delta \alpha/\alpha = -0.574 \pm 0.102 \times
10^{-5}$ [5], but the latest results from another group and telescope
do not see that effect, finding $\Delta \alpha/\alpha = 0.06\pm0.06
\times 10^{-5}$ [6].  This seems to me good news, since such a
variation might be inconsistent with the entire framework of
relativistic quantum field theory [7].  But of course it will be
necessary to test this new result, and to understand the origin of the
earlier apparent variation in $\alpha$.

\section{Cosmological Parameters}

In the mid-1990s there was a crisis in cosmology, because the age of
the old globular cluster stars in the Milky Way, then estimated to be
$16\pm 3$ Gyr, was higher than the expansion age of the universe,
which for a critical density ($\Omega_m=1$) universe is $9\pm2$ Gyr
(using Hubble parameter $h=0.72\pm0.08$).  But when the data from the
Hipparcos astrometric satellite became available in 1997, it showed
that the distances to the globular clusters had been underestimated,
which (combined with improved stellar evolution models) implied that
their ages are $12\pm3$ Gyr.

The successful Hubble telescope key project on the extragalactic
distance scale determined that the Hubble parameter $H_0=100 h$ km
s$^{-1}$ Mpc$^{-1}$ is $h=0.72\pm0.08$ [8].  Several lines of evidence
-- including CBR, supernovae, and clusters -- now show that the
universe does not have $\Omega_m=1$, but rather $\Omega_{tot}=\Omega_m
+ \Omega_\Lambda=1$ with $\Omega_m \approx 0.3$, which gives an
expansion age of about 14 Gyr.  The WMAP cosmic background data alone
give an expansion age of $13.4\pm0.3$ Gyr, which becomes $13.7\pm0.2$
with the WMAP running power spectrum index model [4].

A new type of age measurement based on radioactive decay of
Thorium-232 (half-life 14.1 Gyr) measured in a number of stars gave a
completely independent age of $14\pm 3$ Gyr.  A similar measurement,
based on the first detection in a star of Uranium-238 (half-life 4.47
Gyr), gave $12.5\pm3$ Gyr; a second such star gave an age of
$14.1\pm2.5$ Gyr [9].  These stellar lifetimes are of course lower
limits on the age of the universe.

All the recent measurements of the age of the universe are thus in
excellent agreement.  It is reassuring that three completely different
clocks -- stellar evolution, expansion of the universe, and
radioactive decay -- agree so well.

Ever since the cosmological crisis regarding the age of the universe
was thus resolved, all the data has been consistent with the cosmology
described above, with the main cosmological parameters now all
determined to about 10\% or better [4,11] with the sole exception of
$\sigma_8$, which measures the amplitude of the (linear) power
spectrum on the scale of 8 $h^{-1}$ Mpc.  However, $\sigma_8$ is a
crucial cosmological parameter which has a big influence over the
growth of fluctuations in the early universe.  The current analyses
lead to values of $\sigma_8$ between about 0.7 and 1.1.  But unless
$\sigma_8$ is at least 0.85 or so, it is very hard to see how the
universe could have formed stars and quasars early enough to have
become ionized at $z\sim17$ [12] as indicated by the WMAP detection of
large-angle polarization [13].  The latest analysis of the
cosmological parameters, for the first time including the Lyman
$\alpha$ forest observed in the SDSS quasar spectra along with the
first year WMAP data and the SDSS galaxy clustering data, finds
$\sigma_8=0.90\pm0.03$ and $\Omega_\lambda=0.72\pm0.02$ [11]. This
study finds the primordial spectral index of scalar fluctuations
$n_s=0.98\pm0.02$ with no evidence for running of the spectral index,
equation of state parameter $w\equiv P/\rho = -0.98^{+0.10}_{-0.12}$
at redshift $z=0.3$ with no evidence for variation with redshift, and
a stringent upper limit on neutrino mass $\Sigma m_\nu < 0.42$ eV.
However, the tiny quoted errors do not include systematic
uncertainties in the interpretation of the Lyman $\alpha$ forest data,
which require further analysis.

\section{Possible Problems}

Of course, there are a number of areas in which cosmological theory
and observations are not in obvious agreement.  The space available
does not permit an exhaustive review, so I will concentrate on the
following topics which seem to me to be the most important: big bang
nucleosynthesis, galaxy centers, dark matter substructure, and angular
momentum. Fortunately, some areas which had seemed problematic now
seem less so.  For example, the low values of the quadrupole and
octopole CMB anisotropies reported by WMAP are revised upward in an
improved analysis, and are now in good agreement with the $\Lambda$CDM
predictions [14].  And the possibility that the first year WMAP
data imply that the visible universe might be topologically complex
and smaller than the horizon [15] has now been stringently constrained
by the absence of pairs of circles in the WMAP data [16].  Another
topic that has received more attention in the press than is warranted
by the science is the discovery of massive galaxies at high redshifts
[17].  Although these observations challenge some overly simplified
theories of galaxy formation [18], there are plenty of sufficiently
massive halos at the relevant redshifts (with $\sigma_8=0.9$) to host
the galaxies in question, as Figure 1 shows.  

\begin{figure}[t]  
\centerline{\psfig{file=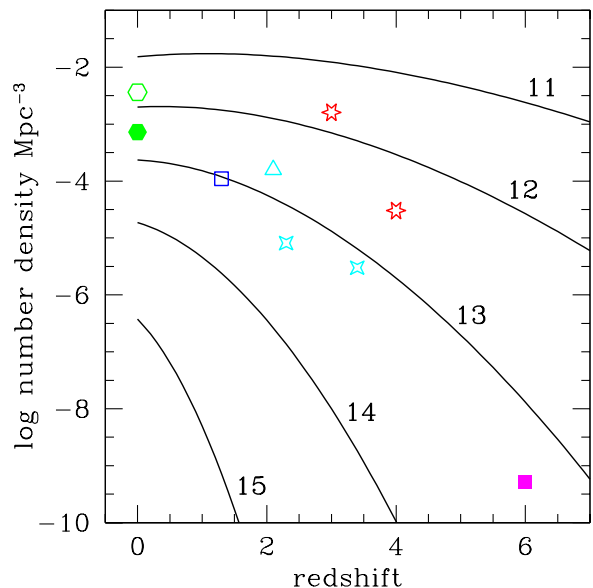, width=3.2in}}
\caption{Curves show the cumulative number density of dark matter
halos more massive than $10^{11}$ to $10^{15} \msun$ (from top to
bottom, as labeled), as a function of redshift, calculated using an
improved Press-Schechter formula [19]. Points show the estimated
comoving number densities of several observed populations, as
follows. Hexagons: galaxies with stellar masses greater than
$10^{11}\msun$ (lower point) and $2.5\times10^{10}\msun$ (higher
point), obtained by integrating the stellar mass function from
SDSS+2MASS; open square: Extremely Red Objects; open triangle: K20
galaxies; crosses: sub-mm galaxies; six pointed stars: Lyman break
galaxies, filled square: quasars. For all populations, $\Lambda$CDM
predicts that there are enough dark matter halos massive enough to
plausibly host the observed objects. (From Ref. [20], which gives the
references for the data points plotted.)}
\end{figure}

\subsection{Big Bang Nucleosynthesis (BBN)}

It is reassuring that the same baryon abundance $\Omega_b h^2 =
0.0214\pm0.0020$ implied by the deuterium abundance D/H in
low-metallicity Lyman limit systems in quasar spectra [21] agrees with
that implied by the relative heights of the first two peaks in the
WMAP angular power spectrum [4], giving $\Omega_b h^2 =
0.0224\pm0.0009$.  This is also in agreement with the baryon abundance
deduced from clusters [22] and with the lower limit from the opacity
of the Lyman alpha forest [23].  However, there are uncertainties in
the measured deuterium abundance evolution [24], and potential
problems -- or perhaps clues to new physics -- in discrepancies
between the observed helium and lithium abundances and the predictions
of BBN.

The abundance by mass of $^4$He measured in low-metallicity ionized
regions in nearby galaxies implies an extrapolated primordial
abundance $Y_p=0.2421\pm0.0021$ according to the latest published data
[25], which corresponds according to standard BBN to a baryon
abundance $\Omega_b h^2 = 0.012^{+0.003}_{-0.002}$, lower by about
$3\sigma$ than the value just mentioned from D/H and CMB measurements.
It remains to be seen whether this could be remedied by improved
analyses (for example, based on more realistic models of these
low-metallicity galaxies or of their HII regions), or alternatively
whether it is perhaps an indication of problems with standard BBN.

The D/H [21] $\Omega_b$ in standard BBN implies a primordial $^7$Li/H
$\approx 3.3-6.0 \times 10^{-10}$, in serious disagreement with the
value of $^7$Li/H $=1.23^{+0.34}_{-0.16} \times 10^{-10}$ measured in
atmospheres of galactic halo metal-poor stars in the ``Spite plateu''
(i.e. with metallicity [Fe/H] less than about -2) [26].  It disagrees
even with the puzzlingly higher value $^7$Li/H $=2.19^{+0.30}_{-0.27}
\times 10^{-10}$ from a sample of globular cluster stars [27].  It is
possible that some of the $^7$Li in such stars is destroyed by
astration, consistent with the small range of $^7$L abundance in the
Spite plateau stars [28], but this may not resolve the discrepancy.

These disagreements call into question the usual assumptions of
standard BBN, for example, the assumption of no significant electron
neutrino asymmetry and $N_\nu=3$ light neutrino species [29].
However, an alternative possibility that might neatly account for the
$^7$Li discrepancy is the injection of energetic nucleons around 1000 s
after the big bang, for example due to decay of the next-to-lightest
supersymmetric partner particle into the lightest one [30].

\subsection{Galaxy and Cluster Centers}

When the first high-resolution simulations of cold dark matter halos
became available [31], they had a central density profile
approximately $\rho(r) \propto r^{-1}$, which has come to be known as
the central ``cusp.''  It was soon pointed out [32,33] that this
central behavior was inconsistent with the HI observations of dwarf
galaxies that were then becoming available, which suggested that the
central density is roughly constant, and also that the first cluster
lensing observations appeared to be inconsistent with a $r^{-\alpha}$
central cusp with $\alpha=1$ [32].  Many additional rotation curves of
low surface brightness (LSB) galaxies were measured, and they also
were claimed to imply that the central density of these galaxies is
rather flat.  It was subsequently realized that the HI observations of
galaxies were affected by finite resolution (``beam smearing''), and
that when this was taken into account the disagreement with
simulations is alleviated [34].  More recently, higher resolution
H$\alpha$ and CO rotation curves have been obtained for a few nearby
dwarf and low surface brightness galaxies [35], and the highest
resolution two-dimensional data imply a variety of central density
profiles ranging from $\alpha\approx 0$ to 1, with evidence for radial
motion especially in the $\alpha \approx 0$ cases [36].

Meanwhile, theorists have done simulations with improving resolution.
On the basis of simulations with tens of thousands of particles per
dark matter halo, Navarro, Frenk, \& White (NFW) [37] showed that
halos from galaxy to cluster scales have density profiles that are
described fairly well by the fitting function $ \rho_{NFW}(r) \equiv
\rho_s (r/r_s)^{-1} (1 + r/r_s)^{-2} $.  Subsequently, James Bullock
[38] and Risa Wechsler [39] improved our understanding of halo
evolution in their dissertation research with me, which included
analyzing thousands of dark matter halos in a high-resolution
dissipationless cosmological simulation by Anatoly Klypin and Andrey
Kravtsov.  Defining the (virial) concentration $ c_{vir} \equiv
R_{vir}/r_s $ (where $R_{vir}$ is the virial radius, Bullock et
al. [38] showed that at fixed halo mass $c_{vir}$ varies with redshift
$z$ as $(1+z)^{-1}$, and developed an approximate mathematical model
that explained the dependence on mass and redshift.  (An alternative
model was proposed in [40], but it appears to be inconsistent with a
recent simulation of small-mass halos [41].)  Wechsler et al. [39]
determined many halo structural merger trees, and showed that the
central scale radius $r_s$ is typically set during the early phase of
a halo's evolution when its mass is growing rapidly, while $c_{vir}$
subsequently grows with $R_{vir}$ during the later slow mass accretion
phase.  Higher resolution simulations with roughly a million particles
per halo gave central density profiles $\rho(r) \propto r^{-\alpha}$
with $\alpha$ as steep as 1.5 [42], although more recent very high
resolution simulations are finding less steep central profiles with
$\alpha \approx 1$ or shallower as $r\longrightarrow 0$ [43].  The
disagreement between the theoretical $\alpha \approx 1-1.5$ vs. the
observed $\alpha \approx 0-1$ may just reflect the effects of baryonic
matter in the centers of galaxies and the difference between circular
velocity and rotation speed likely to arise in gaseous disks embedded
within triaxial halos [44].

The mean density $\Delta_{V/2}$ inside the radius $r_{V/2}$ (where the
rotation velocity reaches half the maximum observed value) appears to
be somewhat smaller than the $\Lambda$CDM prediction with
$\sigma_8=0.9$, but more consistent with $\Lambda$CDM with
$\sigma_8=0.7$ [45] or ``quintessence'' models with equation of state
parameter $w<-1$ [46], but as we have mentioned these possibilities
appear to be inconsistent with other data (e.g. [11]).

In several clusters of galaxies, after removing the baryonic
contribution the central dark matter profile appears to be rather
shallow, with $\alpha \approx 0.35$ for cluster MS2137-23 [46].  The
apparent disagreement with CDM worsens if adiabatic compression of the
dark matter by the infalling baryons is considered [47]. However,
dynamical friction of the dense galaxies moving in the smooth
background of the cluster dark matter counteracts the effect of
adiabatic compression, and can lead to energy transfer from the
galaxies to the dark matter which heats up the central cuspy dark
matter and softens the cusp. N-body simulations [48] show that the
dark matter distribution can become very shallow, with $\alpha \approx
0.3$ for a cluster like MS2137, in agreement with observations.
Taking the triaxial shapes of cluster centers [49] into account may
also help to bring theory and observations into agreement [50].

\subsection{Substructure}

Many fewer small satellite galaxies are seen around the Milky Way and
M31 than the number of small dark matter halos seen in semianalytic
models [51] and high-resolution simulations [52,53] of such systems.
But for $\Lambda$CDM the discrepancy arises only for satellites
smaller than the LMC and SMC [52], and such small satellites are
expected to form stars very inefficiently [54,55].  Semianalytic
models taking both reionization of the universe and local feedback
from supernovae into account appear to be in good agreement with
observations of the relative numbers of faint and bright galaxies in
environments ranging from the local group to clusters [55].  This is
encouraging, but it is important to check whether such models and
high-resolution simulations [56] are capable of accounting for more
detailed properties of low-mass galaxies, such as their radial
distribution, metallicity and ages [57], and the ``fundamental line''
of dwarf galaxy properties [58].

``Millilensing'' by DM halo substructure may be required to account
for anomalous flux ratios in radio lensing of quasars by galaxies
producing multiple images [59]. A concern is whether there enough halo
substructure in the inner $\sim10$ kpc of galaxies, where it appears
to be needed to account for such lensing, although a recent analysis
suggests that substructure along the line of sight to these galaxies
can help account for the observations [60].  Spectroscopic
observations can determine the mass of the lensing perturber by
comparing the magnification of different regions, for example whether
the tiny broad line region nearest the AGN is lensed, or also the much
larger narrow line region [61].  Lensing of AGN jets can also be a
useful diagnostic for substructure [62].  Further observations are
needed.

\subsection{Angular Momentum}

There are two angular momentum problems [63]: (1) overcooling, and (2)
the wrong distribution of angular momentum in halos.  (1) {\it
Overcooling}: For many years, realistic spiral galaxies did not form
in hydrodynamic simulations [64].  However, it is plausible that
unrealistically effective cooling (``overcooling'') in the simulations
was responsible for the loss of angular momentum [65].  More realistic
disk galaxies have formed in recent, higher-resolution simulations
including feedback [66]; an appropriate equation of state for the gas
in galactic disks may play a particularly important role [67].  (2)
{\it Wrong angular momentum distribution}: The standard tidal torque
picture of how dark matter halos and the galaxies that they host get
their angular momentum [68] suggests that the dark matter and baryons
will have similar angular momentum distributions.  The distribution of
the specific angular momentum among the dark matter particles can be
described by a simple fitting function, but disk galaxies like those
seen would not form if the baryons have this same angular momentum
distribution [69].  My colleagues and I have developed an alternative
picture focussing on the angular momentum growth of the largest
progenitor of a given halo, in which the halo's angular momentum comes
mainly from the orbital angular momentum of the accreted halos, and we
showed that this model accurately reproduces simulation results [70].
In this model, large angular momentum changes occur following major
mergers -- but the gas (which shocks) and the dark matter (which does
not) would be expected to behave differently in such mergers.
Steadily improving hydrodynamic simulations are being done to study
these processes in the standard $\Lambda$CDM cosmology [71], and new
techniques are being developed to compare the outputs to observations
[72], but it remains to be seen whether the results will be a good
match to the mostly irregular galaxies observed at high redshift
turning into the observed Hubble sequence of galaxies at low redshift.
Our model of angular momentum growth of dark matter halos implies that
the halos that have not had a recent merger have lower spin parameter
(dimensionless angular momentum) than average.  A perhaps surprising
consequence is that the halos that host elliptical galaxies that
formed from major mergers are expected to have higher angular momentum
than those that host spiral galaxies (since major mergers destroy
galactic disks).  This is contrary to naive expectations [70,73].

\section{Conclusions}

On {\bf large scales}, the agreement between $\Lambda$CDM and
observations is spectacular.  $\Lambda$CDM simulations also account
remarkably well for smaller scale observations such as galaxy
clustering [74], and even for the radial distribution [75] and total
mass associated with galaxies [76].

The problems of {\bf big bang nucleosynthesis} may be a clue to new
particle physics or astrophysics.  On {\bf cusps} there has been
tremendous progress on observing velocity fields in nearby galaxies,
and also real progress in improving simulations. {\it Observed}
simulations may agree with observed velocities in galaxy centers
better than seemed likely a few years ago.  But it is something of a
scandal that there is still so little theoretical understanding of
dark matter halo central behavior, although people are making progress
on this problem [77].  It is likely that triaxial halo structure and
poorly understood gastrophysics will turn out to be relevant.  On dark
matter {\bf halo substructure}, it looked last year as if a challenge
might be turning into a success for CDM, if the amount of substructure
predicted by $\Lambda$CDM is indeed what is required to account for
the number of satellites seen and for the flux anomalies observed in
radio lensing.  The main question is whether the amount of dark matter
in subhalos and the predicted radial distribution of such substructure
agrees with lensing and observed satellites.  Much work remains to be
done to test the theory quantitatively.  Regarding {\bf angular
momentum} problems, resolving the crucial issues again involves
developing better understanding of messy astrophysics.  Fortunately,
in all these areas wonderful new telescopes are providing crucial
data that will help develop and test theory.

At this conference, the updates on direct dark matter detection
experiments showed impressive progress, with greatly expanded
parameter space now being probed experimentally.  Unfortunately, the
relevant weakly interacting massive particle (WIMP) elastic scattering
cross section is quite uncertain.  The annihilation cross section is
much better constrained, since this is what determines the WIMP
abundance today in most models, but the dark matter density at the
center of the Milky Way is uncertain because there are competing
processes that can enhance it or diminish it.  It is possible that the
new H.E.S.S. array of atmospheric Cherenkov telescopes (ACTs) has
discovered dark matter annihilation at the galactic center, with a
dark matter particle mass of approximately 18 TeV [78], which is
unexpectedly high but not impossible for the lightest supersymmetric
partner particle.  The high energy gamma rays appear to come from the
sort of centrally peaked density profiles predicted as a consequence
of scattering of WIMPs by the star cluster surrounding the central
supermassive black hole Sag A$^*$ [79], and the necessary dark matter
density is consistent with theoretical expectations [80] if there is
baryonic contraction [47,81].  It is unlikely that the gamma rays come
from near the black hole event horizon, since there does not appear to
be any time variability in the gamma rays although the X-ray and
optical radiation from the black hole is quite variable.  The main
alternative explanation for the high energy gamma rays is that they
come from the supernova remnant known as Sag A$^*$ East, which covers
a region several pc across surrounding the galactic center.  These
alternatives can perhaps be distinguished via the different angular
distributions expected, when data taken with all four H.E.S.S. ACTs
are analyzed.  Another important discriminant is the energy spectrum
of the gamma rays, which must sharply cut off if the gamma rays come
from annihilation.

\vskip 0.1in
\noindent {\bf Acknowledgments}

I thank David Cline for inviting me to present this opening talk at
Dark Matter 2004, and I thank my colleagues and students -- especially
those who participated in the UCSC Workshop on Galaxy Formation in
August 2004 -- for enlightening me about recent work in cosmology.  I
am grateful to Ben Metcalf for a careful reading of this manuscript.
I acknowledge support from grants NASA NAG5-12326 and NSF AST-0205944.

\def\ApJ{{\sl ApJ}}
\def\apj{{\sl ApJ}}
\def\ApJS{{\sl ApJS}}
\def\MNRAS{{\sl MNRAS}}
\def\mnras{{\sl MNRAS}}
\def\Nature{{\sl Nature}}
\def\AJ{{\sl AJ}}

\newenvironment{reflist}{\begin{list}{}{\listparindent -0.20in
 \leftmargin 0.0in} \item \ \vspace{-.35in} }{\end{list}}

\vskip 0.2in

\begin{reflist}

{\bf References}

\

1. J. R. Primack 2004, IAU Symposium 220 {\it Dark Matter in
Galaxies}, eds. S. D. Ryder et al. (Astron. Soc.  Pacific), p. 53 and
p. 467, and other articles in that volume.  J. R. Primack 2003,
``Status of Cold Dark Matter Cosmology,'' in {\it Proceedings of 5th
International UCLA Symposium on Sources and Detection of Dark Matter},
February 2002, ed. D. Cline, Nucl. Phys. B, Proc. Suppl., {\bf 124},
3 (astro-ph/0205391).

2. S. Weinberg 2001, ``The Non-Revolution of Thomas Kuhn,'' in
   {\it Facing Up: Science and Its Cultural Adversaries} (Harvard
   University Press), p. 205.
   J. R. Primack and N. E. Abrams, ``Scientific Revolutions in
   Cosmology: Overthrowing vs. Encompassing,'' {\it Philosophy in
   Science}, {\bf 9}, 75 (available on the web at
   physics.ucsc.edu/cosmo/primack\_abrams/SciRevolutionsinCosm69B.pdf).
   
3. G. R. Blumenthal, S. M. Faber, J. R. Primack, and M. J. Rees,
   {\sl Nature}, {\bf 311}, 517 (1984). 

4. D. N. Spergel et al. 2003, \ApJS, {\bf 148}, 175.

5. J. K. Webb et al., {\sl Phys. Rev. Lett.}, {\bf 87}, 091301 (2001); 
   M. T. Murphy et al. 2003, \MNRAS, {\bf 345}, 609 (2003).

6. R. Srianand et al. 2004, {\sl Phys. Rev. Lett.}, {\bf 92}, 121302. 
   H. Chand et al. 2004, {\sl A\&A}, {\bf 417}, 853; and astro-ph/0408200.
   Cf. L. L. Cowie and A. Songaila 2004, {\sl Nature}, {\bf 428}, 132.

7. T. Banks, M. Dine, and M. R. Douglas 2002, {\sl Phys. Rev. Lett.},
   {\bf 88}, 131301; M. Dine et al. 2003, {\sl Phys. Rev. D}{\bf 67}, 015009.
   Cf. D. F. Mota and J. D. Barrow 2004, \MNRAS, {\bf 349}, 281.

8. W. L. Freedman et al. 2001, \ApJ, {\bf 553}, 47.

9.  S. Wanajo, N. Itoh, S. Nozawa, and T. C. Beers 2002, \ApJ, 577,
   853.  See the review by J. J. Cowan and C. Sneden 2003, in
{\it Proceedings of the 3rd International Conference on Fission and
Properties of Neturon-Rich Nuclei}, edited by J. H. Hamilton et
al. (World Scientific) (astro-ph/0212149).

10. M. Tegmark et al. 2004, {\sl Phys. Rev. D}{\bf 69}, 103501.

11. U. Seljak et al. 2004, astro-ph/0407372.  Cf. D. Tytler et
    al. 2004, astro-ph/0403688.

12. R. S. Somerville, J. S. Bullock, and M. Livio 2003, \ApJ, {\bf 593}, 616.
B. Ciardi, A. Ferrara, and S. D. M. White 2003, \MNRAS, {\bf344}, 7.

13. A. Kogut et al., \ApJS, 148, 161 (2003).

14. G. Efstathiou, \MNRAS, 348, 885. Cf. A. Slosar, U. Seljak, A.
    Makarov 2004, {\sl Phys Rev D}{\bf 69}, 123003.

15. J.-P. Luminet, J. R. Weeks, A. Riazuelo, R. Lehoucq, \& J.-P. Uzan
2003, \Nature, {\bf 425}, 593.  Cf. J. J. Levin 2002, {\sl
Phys. Reports}, {\bf 365}, 261.

16. N. Cornish, D. N. Spergel, G. D. Starkman, and E. Komatsu 2004,
{\sl Phys. Rev. Lett.}, {\bf 92}, 201302.

17. E.g., G. D. Wirth 2004, \Nature, {\bf 430}, 149; K. Glazebrook et
al. \Nature, {\bf430}, 181; A. Cimatti et al. 2004, \Nature, {\bf430}, 184.

18. Glazebrook et al. [17] chose to compare their observations with
the semianalytic predictions of the Durham group rather than those
of R. S. Somerville, J. R. Primack, and S. M. Faber
2001, \MNRAS, {\bf320}, 504, with which they agree much better (see
[20], which also discusses which observations disagree with
current theoretical galaxy formation models including ours).

19. R. K. Sheth and G. Tormen 1999, \MNRAS, {\bf 308}, 119.

20. R. S. Somerville, in {\it Proceedings of the ESO/USM/MPE Workshop on
  ``Multiwavelength Mapping of Galaxy Formation and Evolution},
    eds. R. Bender and A. Renzini, in press (astro-ph/0401570).

21. D. Kirkman, D.  Tytler, N. Suzuki, J. M. O'Meara, and D. Lubin
    2003, \ApJS, 149, 1.

22. J. E. Carlstrom, G. P. Holder, and E. D. Reese 2002, {\sl ARAA}, 
{\bf40}, 643.

23. M. Rauch et al. 1997, \ApJ, {\bf487}, 7.  Cf. M.  Dijkstra, Mark,
 A. Lidz, and L. Hui 2004, \ApJ, {\bf605}, 7.

24. E.g., N. H. M. Crighton et al. 2004, astro-ph/0403512.

25. Y. I. Izotov and T. X. Thuan 2004, \ApJ, {\bf 602}, 200.

26. S. G. Ryan et al. 2000, \ApJ, {\bf 530}, L57.

27. P. Bonifacio and P. Molaro 1997, \MNRAS, {\bf 285}, 847;
P. Bonifacio, P. Molaro, and L. Pasquini 1997, \MNRAS, 292,  L1.

28. M. H. Pinsonneault et al. 2002, \ApJ, {\bf 574}, 398.

29. J. P. Kneller and G. Steigman, astro-ph/0406320.

30. J. L. Feng, A. Rajaraman, and F. Takayama 2003, {\sl
    Phys. Rev. Lett.}, {\bf 91}, 011302; {\sl Phys. Rev. D}{\bf 68},
    063504.  J. L. Feng, S. Su, and F. Takayama 2004, hep-ph/0404231.
    K. Jedamzik 2004, astro-ph/0402344 and astro-ph/0405583.
    Cf. K. Jedamzik 2000, Phys. Rev. Lett., {\bf 84}, 3248, on $^6$Li.

31. J. Dubinski and R. G. Carlberg 1991, \ApJ, {\bf378}, 496

32. R. Flores and J. R. Primack 1994, \ApJ, {\bf427}, L1.

33. B. Moore 1994, \Nature, {\bf370}, 629.

34. F. C. van den Bosch et al. 2000, \AJ, {\bf119}, 1579; 
F. C. van den Bosch and R. A. Swaters 2001, \MNRAS, {\bf325}, 1017.

35. E.g. W. J. G. de Blok et al. 2001, \apj, 552, L23; 
S. McGaugh et al. 2001, \AJ, {\bf122}, 2381;
R. A. Swaters et al. 2003, \ApJ, {\bf583}, 732;
W. J. G. de Blok et al. 2003, \MNRAS, {\bf340}, 657.

36. A. Bolatto et al. 2002, \ApJ, {\bf 565}, 238; J. D. Simon et
    al. 2003, \ApJ, {\bf 596}, 957; J. D. Simon et al. 2004,
    astro-ph/0310193; A. Bolatto et al. 2004, IAU Symposium 220 {\it Dark
    Matter in Galaxies}, eds. S. D. Ryder et al. (Astron. Soc. Pacific), 
    p. 353.

37. J. F. Navarro, C. S. Frenk, and S. D. M. White 1996, \apj, {\bf462},
563; 1997, \apj, {\bf 490}, 493 (NFW).

38. J. S. Bullock et al. 2001,  \mnras, {\bf321}, 559.

39. R. H. Wechsler, J. S. Bullock, J. R. Primack, A. V. Kravtsov, and
    A. Dekel 2002, \apj, {\bf 568}, 52.  Cf. D. H. Zhao et al. 2003,
    \MNRAS, {\bf 339}, 12; \ApJ, {\bf 597}, L9; and D. Reed et
    al. 2004, astro-ph/0312544.

40. V. R. Eke, J. F. Navarro, and M. Steinmetz 2001, \apj, 554, 114.

41. P. Colin, A. Klypin, O. Valenzuela, and S. Gottloeber 2003,
    astro-ph/0308348.

42. B. Moore et al. 1999,  \mnras, {\bf310}, 1147; S. Ghigna et al. 
2000, \apj, {\bf 544}, 616; A. A. Klypin et al. 2001, \apj, {\bf554}, 903.

43. C. Power et al. 2003, \mnras, {\bf338}, 14; F. Stoehr et al. 2003, 
\mnras, {\bf345}, 1313; J. F. Navarro et al. 2004, \MNRAS, {\bf349}, 1039;
Hayashi 2004, \MNRAS submitted (astro-ph/0310576).

44. J. F. Navarro 2004, IAU Symposium 220 {\it Dark Matter in Galaxies},
eds. S. D. Ryder et al. (Astron. Soc. Pacific), p. 61; E. Hayashi et
al. 2004, astro-ph/0408132.

45. S. M. K. Alam, J. S. Bullock, and D. H. Weinberg 2002, \apj, {\bf 572}, 34.
A. R. Zentner and J. S. Bullock 2002, {\sl Phys Rev. D}{\bf66}, 043003.

46. D. J. Sand, T. Treu, and R. S. Ellis 2002, \ApJ, {\bf 574}, L129; 
T. Treu et al. 2004, IAU Symposium 220 {\it Dark Matter in Galaxies},
eds. S. D. Ryder et al. (Astron. Soc. Pacific), p. 159;
D. J. Sand, T. Treu, G. P. Smith, and R. S. Ellis 2004, \ApJ, {\bf604}, 88.

47. G. R. Blumenthal, S. M. Faber, R. Flores, and J. R. Primack 1986,
\apj, {\bf301}, 27.

48. A. El-Zant, Y. Hoffman, J. R. Primack, F. Combes, and I. Shlosman
2004, \ApJ, {\bf607}, L75; C.-P. Ma and M. Boylan-Kolchin 2004,
{\sl Phys. Rev. Lett.}, {\bf93}, 021301; C. Nipoti, T. Treu, L. Ciotti,
and M. Stiavelli 2004, \MNRAS\  submitted (astro-ph/0404127).

49. J.-P. Jing and Y. Suto 2002, \ApJ, {\bf574}, 538.
R. Flores et al., in prep.

50. M. Bartelmann and M. Meneghetti 2004, {\sl A\&A}, {\bf 418}, 413;
N. Dalal and C. R. Keeton 2003, astro-ph/0312072.

51. G. Kauffmann, S. D. M. White, and B. Guiderdoni 1993, \MNRAS,
{\bf264}, 201.

52. A. A. Klypin et al. 1999, \apj, {\b 522}, 82.

53. B. Moore et al. 1999, \apj, {\bf524}, 19.

54. J. S. Bullock, A. V. Kravtsov, and D. H. Weinberg 2000, \apj,
{\bf539}, 517.

55. R. S. Somerville 2002, \ApJ, {\bf572}, L23; R. B. Tully et
al. 2002, \ApJ, {\bf572}, L23.  A. J. Benson et al. 2003, 
\MNRAS, {\bf343}, 679.

56. A. V. Kravtsov, O. Y. Gnedin, and A. A. Klypin 2004, \ApJ, {\bf609}, 482.

57. B. Willman et al. 2004, \MNRAS\ in press (astro-ph/0403001).
E. K. Grebel and  J. S. Gallagher III 2004, \ApJ, {\bf610}, L89.
C. J. Conselice 2003, {\sl Ap\&SS}, {\bf284}, 631.

58. F. Prada and A.  Burkert 2002, \ApJ, {\bf564}, L73; A. Dekel and J. Woo
2003, \MNRAS, {\bf344}, 1131.

59. R. B. Metcalf and P. Madau 2001, \apj, {\bf 563}, 9.
M. Chiba 2002, \ApJ, {\bf565}, 17.
R. B. Metcalf and H. Zhao 2002, \apj, {\bf567}, L5.
N. Dalal and C. S. Kochanek 2002, \apj, {\bf572}, 25.
P. L. Schechter and J. Wambsganss 2002, \apj, {\bf580}, 685.
C. S. Kochanek and N. Dalal 2004, \ApJ, {\bf610}, 69.
C. R. Keeton et al. 2003, \ApJ, {\bf598}, 138.
S. Mao et al. 2004, \ApJ, {\bf604}, L5. 

60. R. B. Metcalf 2004, astro-ph/0407298.

61. R. B. Metcalf and L. A. Moustakas 2003, \MNRAS, {\bf339}, 607;
R. B. Metcalf et al. 2004, \ApJ, {\bf607}, 43.

62. R. B. Metcalf 2002, \ApJ, {\bf580}, 696.

63. J. R. Primack 2004, IAU Symposium 220 {\it Dark Matter in
Galaxies}, eds. S. D. Ryder et al. (Astron. Soc.  Pacific), p. 467.

64. J. F. Navarro and W. Benz 1991, \apj, {\bf380}, 320;
J. F. Navarro and M. Steinmetz 1997, \apj, {\bf478}, 13.

65. A. H. Maller and A. Dekel 2002, \mnras, {\bf335}, 48.

66. M. G. Abadi et al. 2003, \ApJ, {\bf59}, 499; \ApJ, {\bf597}, 21.
F. Governato et al. 2004, \ApJ, {\bf607}, 688.  

67. B. Robertson et al. 2004, \ApJ, {\bf606}, 32.

68. P. J. E. Peebles 1969, \ApJ, {\bf155}, 393; S. D. M. White 1984, \ApJ,
{\bf 286}, 38.

69. J. S. Bullock et al. 2001, \ApJ, {\bf555}, 240.

70. M. Vitvitska et al.  2002, \ApJ, {\bf581}, 799;
A. H. Maller, A. Dekel, and R. S. Somerville 2002, \mnras, {\bf329}, 423.

71. F. C. van den Bosch et al. 2003, \ApJ, {\bf576}, 21; \MNRAS, {\bf346}, 177.
D. N. Chen, Y.-P. Jing, and K. Yoshikawa 2003, \apj, {\bf597}, 35.
J. Barnes 2002, \MNRAS, {\bf333}, 481.
T. J. Cox, J. R. Primack, P. Jonsson, and R. S. Somerville 2004, ApJ,
607, L87.  T. J. Cox 2004, UCSC PhD dissertation.

72. C. J. Conselice et al. 2004, 2004, \ApJ, {\bf600}, L139.
P. Jonsson 2004, UCSC PhD dissertation.
J. Lotz, J. R. Primack, and P. Madau 2004, \AJ, {\bf128}, 163.

73. E. D'Onghia and A. Burkert 2004, \ApJ, {\bf612}, L13.

74. E.g. A. V. Kravtsov et al. 2004, \ApJ, {\bf609}, 35.

75. F. Prada et al. 2003, \ApJ, {\bf598}, 260.

76. A. Tasitsiomi et al. 2004, astro-ph/0404168.

77. E.g. Dekel et al. 2003, \apj, {\bf588}, 680; \mnras, {\bf341}, 326.
L. L. R. Williams, A. Babul, and J. J. Dalcanton 2004, \ApJ, {\bf604}, 18.

78. F. Aharonian et al. 2004, {\sl A\&A} in press (astro-ph/0408145);
D. Horns 2004, astro-ph/0408192.  Cf. R. Irion 2004, {\sl Science}, {\bf
305}, 763.

79. O. Y. Gnedin and J. R. Primack 2004, {\sl Phys. Rev. Lett.} in press
    (astro-ph/0308385); D. Merritt 2004, {\sl Phys. Rev. Lett.}, {\bf92},
    201304. 

80. F. Prada et al. 2004, astro-ph/0401512.

81. O. Y. Gnedin et al. 2004, astro-ph/0406247.
\end{reflist}
\end{document}